\documentclass[11pt]{article}
\usepackage{epsf,amssymb,cite}
\newcommand{\xtra}[1]{{.}}
\renewcommand{\xtra}[1]{{, \tt hep-th/#1.}}

\newcommand{\cR}{{\cal R}}
\newcommand{\One}{\hbox{{\rm 1{\hbox to 1.5pt{\hss\rm1}}}}}
\newcommand{\Onesmall}{\hbox{$\scriptstyle 1${\hbox to .5pt{\hss$\scriptstyle
1$}}}}

\newcommand{\ri}{\right}
\newcommand{\lf}{\left}

\newcommand{\Ga}{\Gamma}

\newcommand{\eq}{\begin{equation}}
\newcommand{\en}{\end{equation}}
\newcommand{\bea}{\begin{eqnarray}}
\newcommand{\eea}{\end{eqnarray}}

\newcommand{\ba}{\begin{array}}
\newcommand{\ea}{\end{array}}

\newcommand{\CC}{{\hbox{\rm C\kern-0.5em{$\sf I$}}}}
\newcommand{\II}{\hbox{{\rm l{\hbox to 1.5pt{\hss\rm l}}}}}
\newcommand{\RR}{{\hbox{$\rm\textstyle I\kern-0.2em R$}}}
\newcommand{\ZZ}{{\hbox{$\sf\textstyle Z\kern-0.4em Z$}}}

\newcommand{\resection}[1]{\setcounter{equation}{0}\section{#1}}

\newcommand{\NP}[1]{Nucl.\ Phys.\ {\bf #1}}
\newcommand{\PL}[1]{Phys.\ Lett.\ {\bf #1}}
\newcommand{\CMP}[1]{Commun.\ Math.\ Phys.\ {\bf #1}}

\newcommand{\PTPS}[1]{Prog.\ Theor.\ Phys.\ Suppl.\ {\bf #1}}
\newcommand{\MPL}[1]{Mod.\ Phys.\ Lett.\ {\bf #1}}
\newcommand{\IJMP}[1]{Int.\ J.\ Mod.\ Phys.\ {\bf #1}}

\newcommand{\TMP}[1]{Teor.\ Math.\ Phys.\ {\bf #1}}
\newcommand{\AlBZ}{Al.B.~Zamolodchikov}

\newcommand{\JP}[1]{J.\ Phys.\ {\bf #1}}

\newcommand{\usbl}[1]{\left(#1\right)}
\newcommand{\inti}{\int^{\infty}_{-\infty}}

\newcommand{\fract}[2]{{\textstyle\frac{#1}{#2}}}

\newcommand{\opnup}[1]{\renewcommand{\\}{\\[50 pt]}}
\renewcommand{\bar}{\overline}

\newcommand\im{\hbox{Im}\,}

\newcommand\LYM{{\cal M}(2/5)}

\renewcommand\hat{\widehat}
\begin{document}
%
\begin{titlepage}
\vskip 0.5cm
\begin{flushright}
DTP-98/71 \\
KCL-MTH/98-40 \\
T-98/106 \\
{\tt hep-th/9810098}\\
October 1998 
\end{flushright}
\vskip 1.2cm
\begin{center}
{\Large {\bf Generalisations of the Coleman-Thun mechanism}} \\[5pt]
{\Large {\bf and boundary reflection factors } }
\end{center}
\vskip 0.8cm
\centerline{Patrick Dorey%
\footnote{e-mail: {\tt P.E.Dorey@durham.ac.uk}},
Roberto Tateo\footnote{e-mail: {\tt Tateo@wasa.saclay.cea.fr}}
and G\'erard Watts\footnote{e-mail: {\tt gmtw@mth.kcl.ac.uk}}
}
\vskip 0.6cm
\centerline{${}^1$\sl Department of Mathematical Sciences,}
\centerline{\sl  University of Durham, Durham DH1 3LE, 
England\,}
\vskip 0.2cm
\centerline{${}^2$\sl Service de Physique Th\'eorique, CEA-Saclay,}
\centerline{\sl F-91191 Gif-sur-Yvette Cedex, France\,}
\vskip 0.2cm
\centerline{${}^3$\sl Mathematics Department, }
\centerline{\sl King's College London, Strand, London WC2R 2LS, U.K.}
\vskip 0.9cm
\begin{abstract}
\vskip0.15cm
\noindent
We make a complete pole analysis of the reflection factors of the 
boundary scaling Lee-Yang model. In the process we uncover a number
of previously unremarked mechanisms for the generation of simple
poles in boundary reflection factors, which have implications for
attempts to close the boundary bootstrap in more general models.
We also explain how different boundary conditions can sometimes 
share the same fundamental reflection factor, by relating the phenomenon
to potential ambiguities in the interpretation of certain poles. In the
case discussed, this ambiguity can be lifted by specifying the sign of
a bulk-boundary coupling.
While the recipe we employ for the association of poles with general
on-shell diagrams is empirically correct, we stress that a justification 
on the basis of more fundamental principles remains a challenge for future
work.
\end{abstract}
\end{titlepage}
\setcounter{footnote}{0}
\def\thefootnote{\fnsymbol{footnote}}

\resection{Introduction}
The paper by Ghoshal and Zamolodchikov~\cite{GZa} is responsible for
a fair amount of the recent
work on integrable quantum field theories in the presence of
boundaries. The setup they discussed involves a massive model
defined on the half line $x\in(-\infty,0]$, with an integrable 
boundary condition imposed at $x=0$. The bulk S-matrix must
then be supplemented by a boundary S-matrix (a set of so-called
reflection factors) describing how each
particle bounces off the boundary. This boundary S-matrix is
constrained by various consistency conditions -- the boundary
Yang-Baxter~\cite{Ca}, bootstrap~\cite{FKa,GZa},
unitarity and cross-unitarity~\cite{GZa} equations --
and one aspect of boundary integrability has been the attempt
to match abstract solutions of these constraints with definite boundary
conditions for particular models. Even in relatively simple Lagrangian
field theories such as the affine Toda models~\cite{FKa,Sa,CDRSa}, 
this task has proved to
be rather hard, a major difficulty being the greatly increased
complexity of perturbation theory once spatial translational symmetry
has been broken by the presence of the 
boundary~\cite{CDRSa,CDRa,Ka,Cb,Ta,PBa}. 
As an alternative line
of attack, one might think to check first of all the internal
field-theoretic
coherence of any proposed boundary S-matrix,
before going on to match it with a specific boundary
condition. In particular, it is possible to ask whether its poles
are telling a sensible story. Ever since the work of
Coleman and Thun~\cite{CTa}, the rules for playing this game
in the bulk have been well established. However, in boundary situations
the stock of reliable examples is much more limited, and the
situation has remained relatively unclear. 

The purpose of this note is to analyse a
case which, though simple, seems to exhibit most phenomena one could
hope to meet in more complicated situations. This is the boundary
scaling Lee-Yang model, which was studied from various points of
view in ref.~\cite{Us1}. A combination of TBA and TCSA
techniques, described in~\cite{Us1}, allowed the boundary
S-matrices to be pinned down unambiguously, and these form the
starting-point for the current investigation. We complete the boundary
bootstrap, and find `prosaic' (ie, potentially field-theoretic)
explanations for all physical-strip poles. In some situations,
physical-strip zeroes of reflection factors have a role to play,
providing the boundary analogue of a mechanism previously observed in
the bulk behaviour of the 
non-self-dual affine Toda field theories~\cite{CDSa}.

Earlier work on the closure of the boundary bootstrap includes the papers
by Corrigan et al~\cite{CDRSa}, Fring and K\"oberle~\cite{FKb}, and 
Saleur and Skorik~\cite{SSa}.

\resection{The models and their reflection factors}
For a more leisurely discussion of the boundary scaling Lee-Yang models,
the reader is referred to ref.~\cite{Us1}\,; here we just summarise the
relevant details. In the bulk, the models are perturbations of the
$\LYM$ minimal model by its unique relevant spinless field $\varphi$\,;
at the boundary, the possibilities for
perturbation depend on which of the two possible conformal boundary
conditions is present in the unperturbed model. One, denoted
$\One$ in \cite{Us1}, does not have any relevant boundary fields, while
the other, denoted $\Phi$ in \cite{Us1}, has a single relevant boundary
field $\phi$ with scaling dimension $x_{\phi}=-1/5$. In the latter
case the general perturbed action is thus
\eq
{\cal A}_{\lambda,\Phi(h)}={\cal A}_{\Phi}+
\lambda\!\inti\!dy\int_{-\infty}^0\!dx\,\varphi(x,y)+
h\!\inti\!dy\,\phi(y)~,
\label{arnie}
\en
where ${\cal A}_{\Phi}$ denotes an action for $\LYM$ with the
$\Phi$ conformal boundary condition
imposed at $x=0$, and the bulk and boundary couplings are
$\lambda$ and $h$ respectively. The action ${\cal A}_{\lambda,\One}$ is
similar, but lacks the final term on the right-hand side. For
$\lambda>0$ (in the normalisations of~\cite{Us1}) the bulk behaviour is
in all cases
described by a massive scattering theory with just one particle type,
and S-matrix~\cite{CMa}
\eq
S(\theta)=-\usbl{1}\usbl{2}~~,\quad
\usbl{x}={\sinh\bigl({\theta\over 2}+{i\pi x\over 6}\bigr)\over
        \sinh\bigl({\theta\over 2}-{i\pi x\over 6}\bigr)}~.
\label{asm}
\en
The mass $M$ of this particle is related to the bulk coupling $\lambda$
as $M(\lambda)=\kappa\lambda^{5/12}$, where~\cite{Zg}
$\kappa=2^{19/12}\sqrt{\pi}\lf(\Ga(3/5)\Ga(4/5)\ri)^{5/12}%
\!/\!\lf( 5^{5/16}\Ga(2/3)\Ga(5/6)\ri)$.
There is a single $\phi^3$-style coupling $f$, seen in the residue
of $S(\theta)$ at $\theta=2\pi i/3\equiv iU$, 
the position of the forward-channel pole:
\eq
S(\theta)\simeq \frac{if^2}{\theta-i U}\,,
\qquad U=\frac{2\pi}{3}\,,\qquad f=i\,2^{1/2}3^{1/4}\,.
\label{fsign}
\en
The sign of $f$ is tied to the normalisation of the
scattering states, so here we are implicitly settling on one
particular choice.
At the boundary, reflection factors were proposed and checked against
finite-size data in ref.~\cite{Us1}. For the $\One$ and $\Phi(h)$
boundary conditions,
they can be written in terms of the blocks $\usbl{x}$ defined in
(\ref{asm}) as
\eq
R_{(1)}= 
\usbl{\fract{1}{2}}\usbl{\fract{3}{2}}\usbl{\fract{4}{2}}^{-1}
\en
and
\eq
R_b=
\usbl{\fract{1}{2}}
\usbl{\fract{3}{2}}
\usbl{\fract{4}{2}}^{-1}\!
\usbl{\fract{1-b}{2}}^{-1}\!
\usbl{\fract{1+b}{2}}
\usbl{\fract{5-b}{2}}
\usbl{\fract{5+b}{2}}^{-1}
\label{Rbdef}
\en
respectively. The presence of a boundary scale at the $\Phi(h)$
boundary accounts for the dependence of its reflection factor on the
parameter $b$, which is related to the bulk and boundary couplings
via~\cite{Us1,Us2} 
\eq
h(b)=
\sin\!\lf((b{+}0.5)\frac{\pi}{5}\ri)h_{\rm crit}\,,~~
h_{\rm crit}=
{-}\frac{5^{1/4}2^{4/5}\pi^{3/5}\sin(2\pi/5)}%
{\lf(\Gamma(3/5)\Gamma(4/5)\ri)^{1/2}}%
\lf(\frac{\Gamma(2/3)}{\Gamma(1/6)}\ri)^{\!6/5}%
\!M(\lambda)^{6/5}.
\label{fozzie}
\en

\resection{Pole analysis of the basic reflection factors}

The $R_{(1)}$ reflection factor
was discussed by Ghoshal and Zamolodchikov~\cite{GZa}, albeit
without reference to any specific boundary condition.
The only details we 
can add are the previously-mentioned association of $R_{(1)}$
with the $\One$ boundary, and
the observation that finite-size data shows no evidence of
boundary bound states in this case~\cite{Us1,Us3}. This
is in accord with the assumptions of ref.~\cite{GZa}, where 
an explanation for the physical-strip poles in $R_{(1)}$
was sought involving just 
bulk particles and the `boundary ground state', which we
shall label $\One_0$. (Later, we shall show that an alternative
interpretation of $R_{(1)}$ is also possible, but this is not 
relevant for the $\One$ boundary
condition.)
For reflection factors, the physical strip is
the region
$0\le\im\theta\le\pi/2$ of the complex $\theta$-plane, so $R_{(1)}$
has two physical strip poles, one at $i\pi/6$ 
and one at $i\pi/2$. Following~\cite{GZa}, both can be explained
by postulating 
the existence of a `boundary-particle coupling' $g_{\Onesmall_0}$.
The relevant on-shell diagrams are shown in figures la and 1b.
{\begin{figure}[ht]
\[\begin{array}{cc}
\epsfysize=.28\linewidth\epsfbox{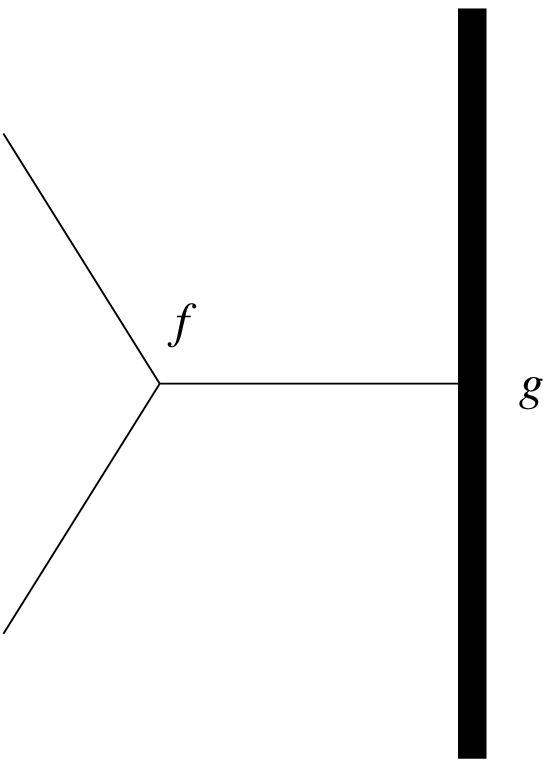}
{}~~~~~~&
\epsfysize=.28\linewidth\epsfbox{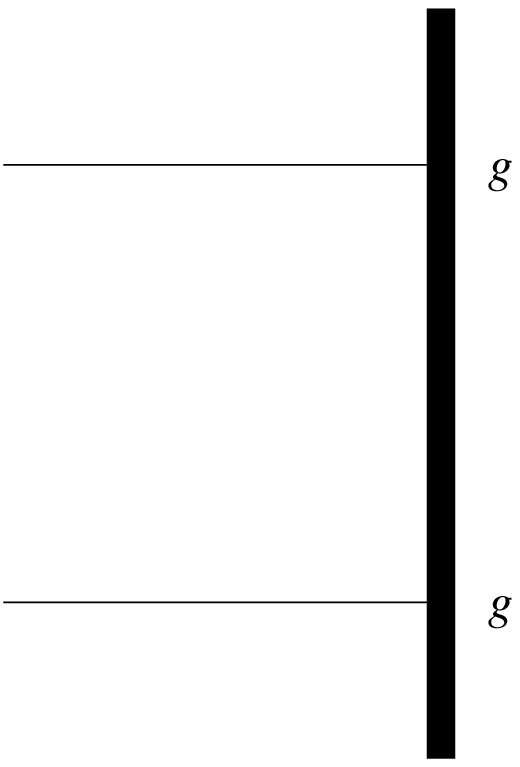}~~~{}
\\[2pt]
\parbox[t]{.4\linewidth}{\small 1a)
The pole in $R_{(1)}(\theta)$ at $i\pi/6$}
{}~&~
\parbox[t]{.4\linewidth}{\small 1b)
The pole in $R_{(1)}(\theta)$ at $i\pi/2$}
\end{array}\]
\end{figure}}

\noindent
A value can be assigned to the coupling $g_{\Onesmall_0}$ using one
of the rules given in~\cite{GZa}. This determines
$g_{\Onesmall_0}$ through the residue of
the pole in $R_{(1)}$ at $\theta=i\pi/6=i\bar U/2$, where
$\bar U=\pi{-}U$ and $U$ is the bulk fusing angle~(\ref{fsign}):
\eq
R_{(1)}(\theta)\simeq \frac{i\,f\,g_{\Onesmall_0}}{2\theta-i\bar U}\,,
\qquad \bar U=\frac{\pi}{3}\,,
\qquad g_{\Onesmall_0}=-i\,2\sqrt{2\sqrt{3}{-}3}\,.
\en
Note that there is no ambiguity about the sign of $g_{\Onesmall_0}$,
once that of $f$ has been fixed.
Alternatively, the square of the coupling was related in~\cite{GZa} to
the residue at $i\pi/2$, via
$R_{(1)}(\theta)\simeq
i(g_{\Onesmall_0})^2/(2\theta{-}i\pi)$. These can be considered as
two definitions of the on-shell coupling, but it must be checked that
they are consistent.
In fact, this is a consequence of the bootstrap and crossing. To see
this, first define, as in ref.~\cite{DTb}, $\cR F(\theta)$ to be equal
to the leading coefficient of the Laurent expansion of the function 
$F$ about
$\theta$ (for simple poles, it is just the residue). In this notation,
the first equation in (\ref{fsign}) becomes
$f^2=-i\cR S(2\pi i/3)$, 
and $(g_{\Onesmall_0})^2$ can either be evaluated as 
$-(2\cR R_{(1)}(i\pi/6)/f)^2$, 
or as $-2i\cR R_{(1)}(i\pi/2)$. The
equality of these two follows from the identity
\eq
\frac{R_{(1)}(\frac{i\pi}{6}{+}\theta)R_{(1)}(\frac{i\pi}{6}{-}\theta)}%
{S(\frac{2\pi i}{3}{-}2\theta)} =
R_{(1)}({\textstyle\frac{i\pi}{2}}{+}\theta)
\label{rident}
\en
which is itself a consequence of the boundary bootstrap equation,
together with unitarity, crossing and crossing-unitarity.

The situation for $R_b$ is richer. We will initially consider
cases where $b$ is real, and can immediately use the symmetries of
(\ref{Rbdef}) to restrict $b$ to the range $[-3,3]$. Figure~2 shows
how the positions of the poles and zeroes
of $R_b$ (which all lie on the imaginary $\theta$-axis for $b$ real)
depend on $b$.
{\begin{figure}[ht]
\[\begin{array}{lc}
\epsfxsize=.38\linewidth\epsfbox{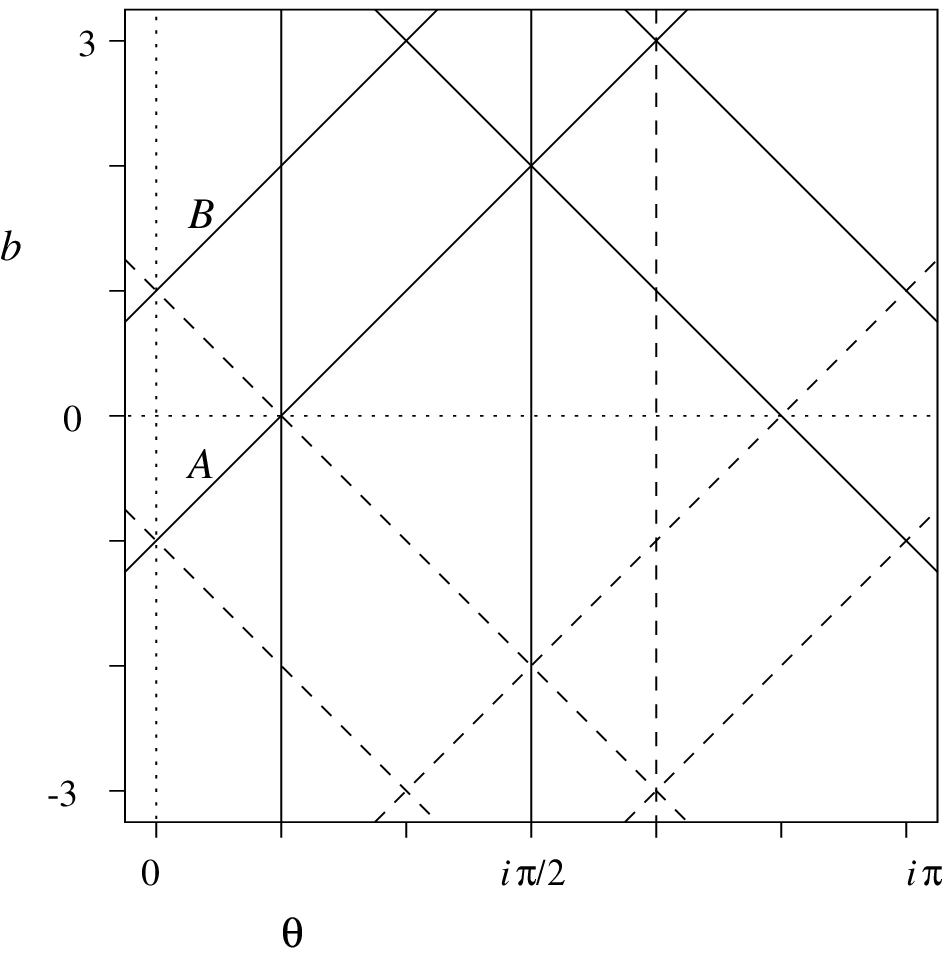}
{}~~~~~~&~~~
\raisebox{24pt}{\epsfysize=0.28\linewidth\epsfbox{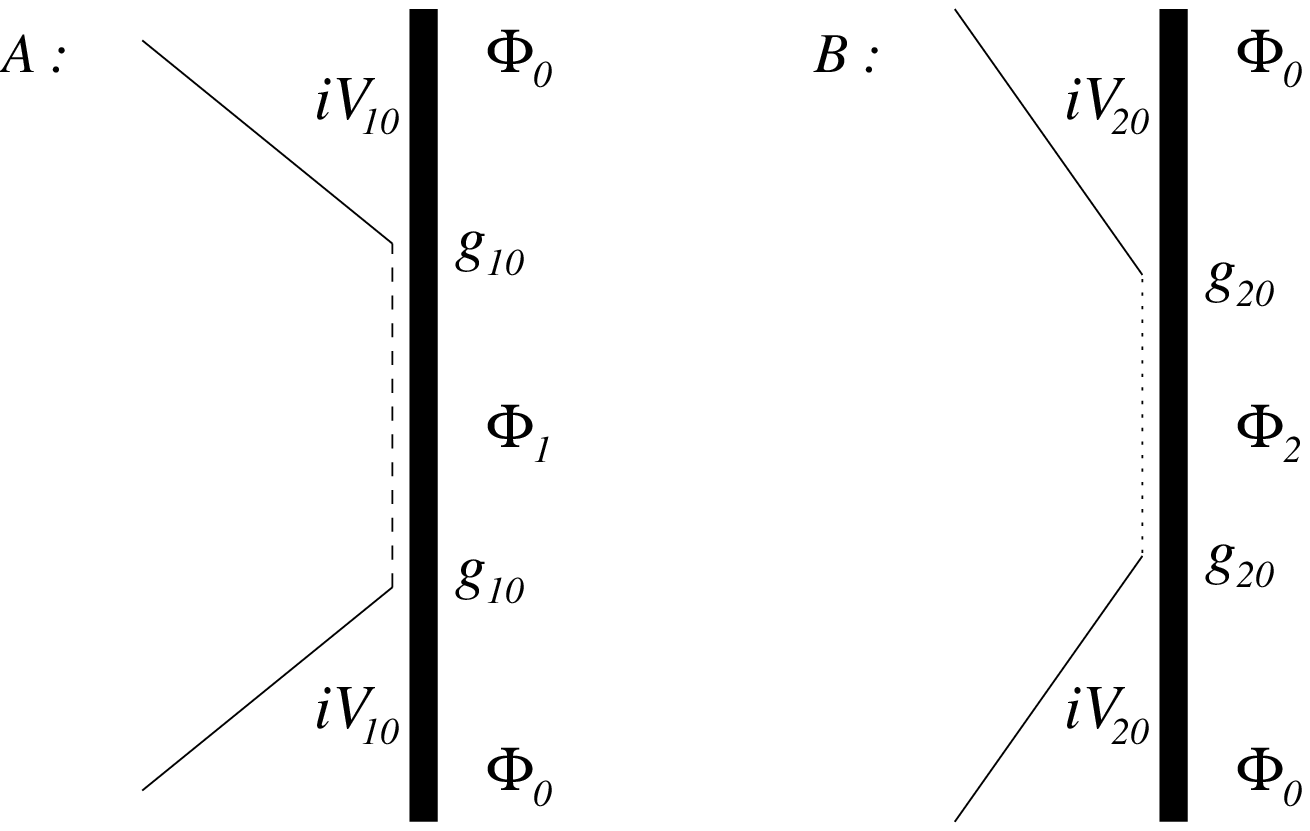}~~~{}}
\\[2pt]
\parbox[t]{.43\linewidth}{\small 2)
Poles and zeroes of $R_b(\theta)\,$. Continuous lines are poles, dashed
lines zeroes, and the dotted lines are the $\theta$ and $b$
axes.}
{}~~&~~
\parbox[t]{.47\linewidth}{\small 3)
The poles $A$ and $B$ in $R_b(\theta)$, at $iV_{10}$ and
$iV_{20}\,$. $\Phi_1$ is indicated by a
dashed line along the boundary, and $\Phi_2$ by a dotted line.}
\end{array}\]
\end{figure}}

Note that, while the relation
(\ref{fozzie}) between $b$ and $h$ has a symmetry about
the point $b{=}2$, the same is clearly not true of the reflection
factor $R_b$. This issue was discussed in the context of the boundary
TBA (BTBA) in ref.~\cite{Us1}, where it was found that the BTBA equations
rearrange themselves as the point $b{=}2$ is passed, so that $R_b$ is
replaced by $R_{4{-}b}$ and the solutions to the equations are indeed
symmetrical about $b{=}2$. In spite of this fact, it turns out that
$R_b$ does continue to have an interpretation for $b>2$.
As found in
ref.~\cite{Us1}\,, the value
of the boundary coupling at $b{=}2$, namely $h_{\rm crit}$, is the
`critical' value at which the energy of the first excited state
(a boundary bound state) becomes degenerate with that of the vacuum.
It is therefore natural to expect that continuation through $b{=}2$
implements a swapping of the vacuum boundary state and the first boundary
bound state, and it will be seen shortly that this is precisely what
happens at the level of the reflection factors, with $R_{b>2}$
becoming the reflection factor for the first boundary bound state. 

First, though, we will treat the pole structure while $b$ remains
in the
range $[-3,2]$. Apart from the exceptional points $b=\pm 2$, there are
always simple poles at $\theta=i\pi/6$ and $\theta=i\pi/2$. These are
explained, just as previously, by postulating a boundary-particle coupling
$g_{\Phi_0}$ between the bulk particle and $\Phi_0$, the boundary
ground state for the $\Phi(h)$ boundary condition. For brevity, we
will write this coupling as
$g_0$. Its value
depends on $h$, or equivalently on $b$, in the following way:
\eq
g_0(h(b))=
\frac{\tan((b{+}2)\pi/12)}{\tan((b{-}2)\pi/12)}~g_{\Onesmall_0}\,.
\label{ghform}
\en
The sequence of
arguments culminating in equation (\ref{rident}) applies here
essentially unchanged, and so there is no need to make a separate
check of the compatibility of the residues at 
$i\pi/6$ and $i\pi/2$. 
(A small subtlety arises when
$b{=}0$, but this will be described later.)

For $b\in [-3,-1]$, this is the end of the story. Then as $b$ passes $-1$, 
an extra pole enters the physical strip, at $i\pi(b{+}1)/6$\,; and
as
$b$ passes $1$, a second appears, at $i\pi(b{-}1)/6$\,. These
$b$-dependent poles are labelled $A$ and $B$ in figure~2. It is natural
to associate them with boundary bound states, and an analysis of
finite-size data confirms this~\cite{Us1}. Writing the two
states as $\Phi_1$ and $\Phi_2$, the situation can be formalised by
introducing a couple of further boundary-particle couplings
$g_{10}$ and 
$g_{20}$, with corresponding `boundary fusing angles' 
$V_{10}=(b{+}1)\pi/6$ and 
$V_{20}=(b{-}1)\pi/6$, as depicted in figure~3. 
The energies $e_1$ and $e_2$
of the boundary bound states are given by
$e_j{-}e_0=M\cos(V_{j0})$,
where $e_0$ is the energy of $\Phi_0$. (As mentioned above, 
$e_1{-}e_0=0$ at $b{=}2$.) Finally, the reflection factors $R^{[1]}_b$
and $R^{[2]}_b$ for a
particle bouncing off the two boundary bound states can be computed,
using the `boundary bound-state bootstrap equation'~\cite{GZa}\,:
\eq
R^{[j]}_b(\theta)=S(\theta{-}iV_{j0})
S(\theta{+}iV_{j0})R_b(\theta)~,\qquad j=1,2.
\en
This yields
\bea
R^{[1]}_b&=&
\usbl{\fract{1}{2}}
\usbl{\fract{3}{2}}
\usbl{\fract{4}{2}}^{-1}\!
\usbl{\fract{1+b}{2}}
\usbl{\fract{3-b}{2}}
\usbl{\fract{3+b}{2}}
\usbl{\fract{5-b}{2}}\,,
\label{Rbidef}\\[3pt]
R^{[2]}_b&=&
\usbl{\fract{1}{2}}
\usbl{\fract{3}{2}}
\usbl{\fract{4}{2}}^{-1}\!
\usbl{\fract{1-b}{2}}^{-1}\!
\usbl{\fract{1+b}{2}}^2\!
\usbl{\fract{3-b}{2}}
\usbl{\fract{3+b}{2}}
\usbl{\fract{5-b}{2}}^2\!
\usbl{\fract{5+b}{2}}^{-1}\,.
\label{Rbiidef}
\eea
The exchange of the vacuum and the
first boundary bound state as $b$ passes through $2$
can now be seen in the equality $R^{[1]}_{4-b}(\theta)=R_b(\theta)$,
whilst the symmetry
$R^{[2]}_{4-b}(\theta)=R^{[2]}_b(\theta)$ reflects
the absence of any such exchange 
at $b{=}2$
involving the second boundary bound state.
The functions $R^{[1]}_b$ and $R^{[2]}_b$ are shown
in figures 4 and 5, and the next section is concerned with their pole
structures.
{\begin{figure}[ht]
\[\begin{array}{cc}
\epsfxsize=.38\linewidth\epsfbox{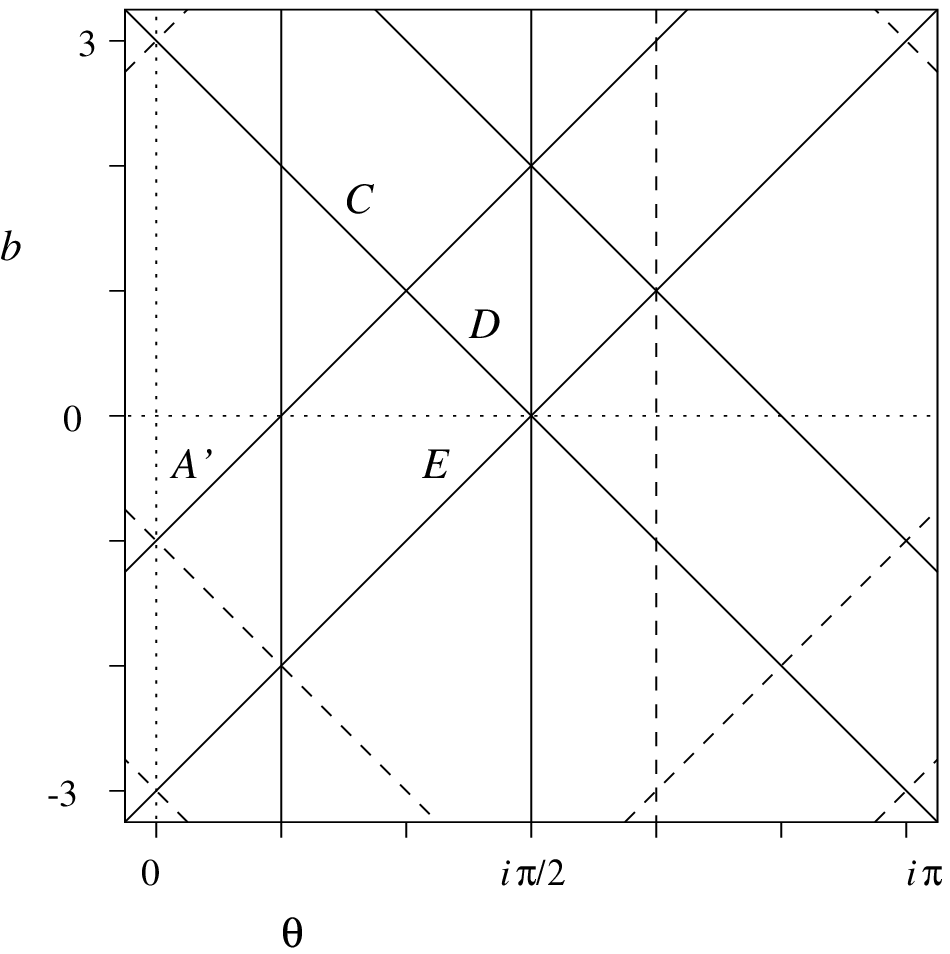}
{}~~~~~~&~~~
\epsfysize=0.38\linewidth\epsfbox{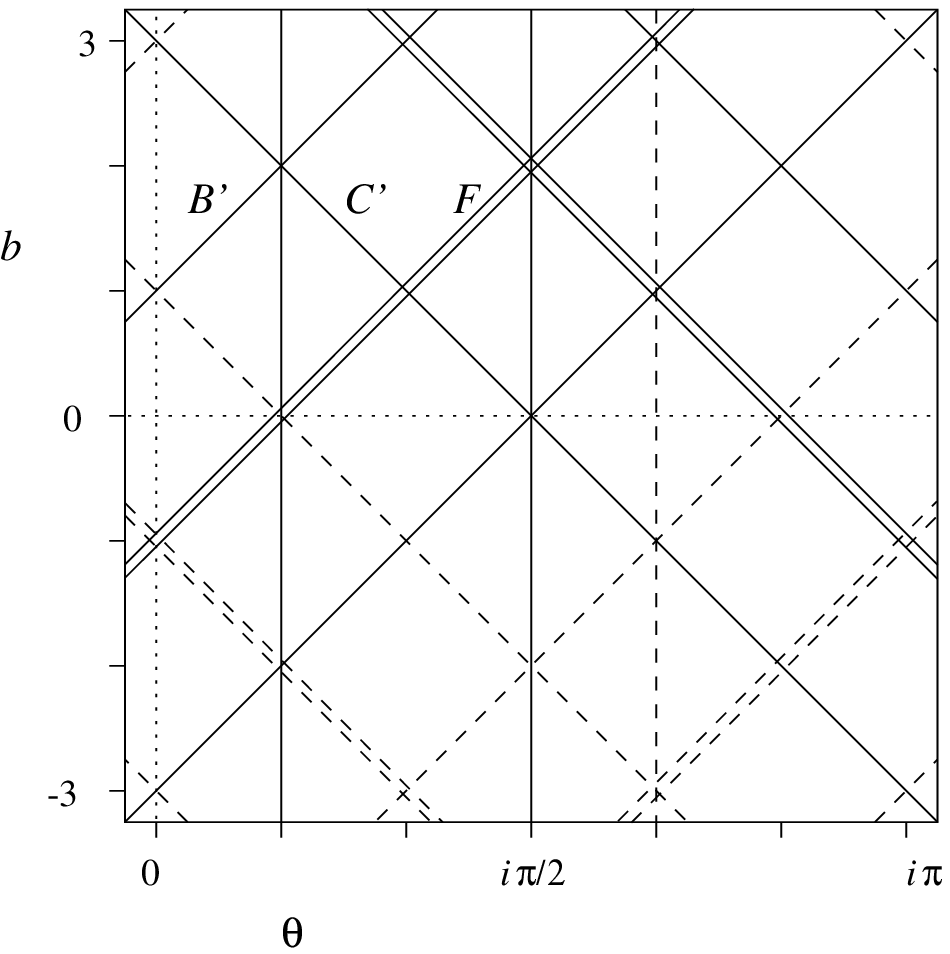}~~~~~{}
\\[2pt]
\parbox[t]{.43\linewidth}{\small 4)
Poles and zeroes of $R^{[1]}_b(\theta)$. Notation as in figure~2.}
{}~~&~~
\parbox[t]{.43\linewidth}{\small 5)
Poles and zeroes of $R^{[2]}_b(\theta)$. Notation as in figure~2, with
double poles and zeroes indicated by double lines.}
\end{array}\]
\end{figure}}

\resection{Pole analysis of the higher reflection factors}
The treatment of the higher reflection factors can be confined to
those ranges of $b$ for which the relevant boundary bound state is
present in the spectrum of the model. For $R_b^{[1]}$ this is the
range $[-1,2]$, and for $R_b^{[2]}$, the range $[1,2]$. The poles in
both reflection factors at $i\pi/6$ and $i\pi/2$ are dealt with as
before, via diagrams of the sort shown in figure~1.

{\begin{figure}[ht]
\[\begin{array}{lc}
{\epsfysize=0.28\linewidth\epsfbox{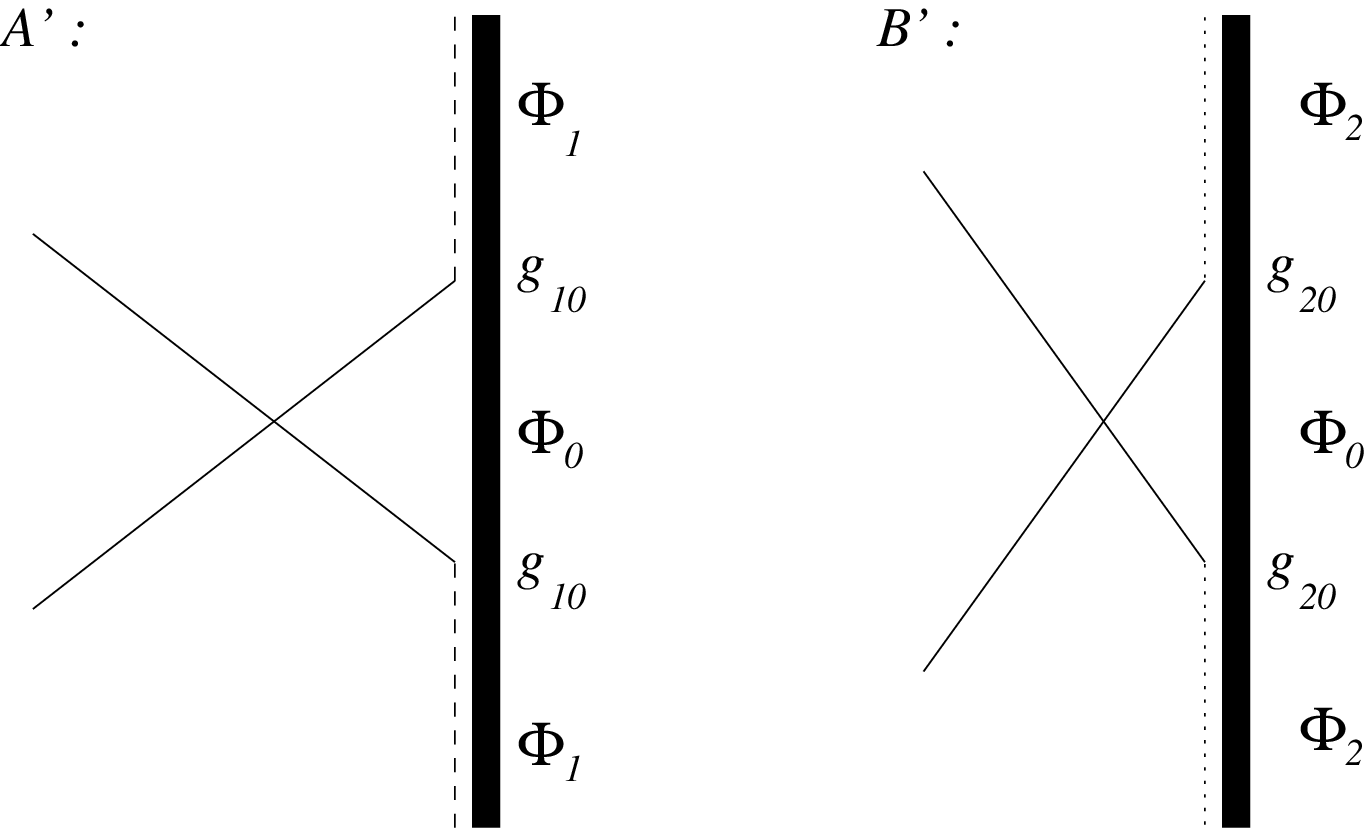}}
{}~~~~~~&~~~
{\epsfysize=0.28\linewidth\epsfbox{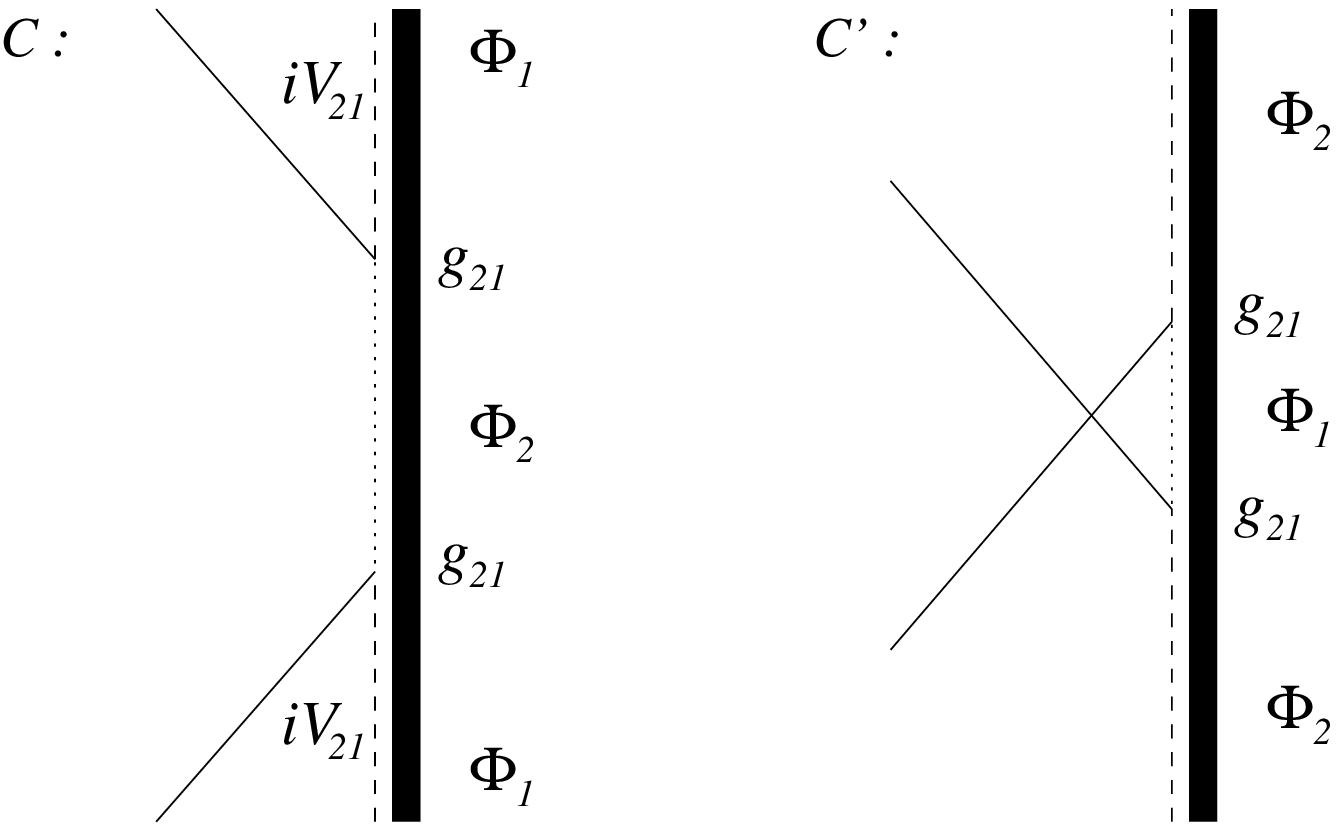}~~~{}}
\\[2pt]
\parbox[t]{.47\linewidth}{\small 6)
The poles $A'$ in $R^{[1]}_b(\theta)$ and $B'$ in $R^{[2]}_b(\theta)$,
 at $iV_{10}$ and
$iV_{20}$ respectively. Notation and fusing angles as in figure~3.}
{}~~&~~
\parbox[t]{.47\linewidth}{\small 7)
The poles $C$ in $R^{[1]}_b(\theta)$ and $C'$ in $R^{[2]}_b(\theta)$,
both at $iV_{21}$.
Notation as in figure~3.}
\end{array}\]
\end{figure}}
 
Next, consider the pole in $R^{[1]}_b$ at $iV_{10}$, labelled 
$A'$ in figure~4. This occurs at the same place as the pole in the
basic reflection factor $R_b$ which
gave rise to $R^{[1]}_b$ in the first place. 
Its existence can be traced to the first on-shell diagram of
figure~6, the boundary analogue of a $u$-channel diagram in the bulk.
An identical mechanism accounts for the pole $B'$ in $R^{[2]}_b$, and
the relevant on-shell diagram is also shown.  

This `$u$-channel' mechanism is a generic feature of
higher reflection factors, and does not seem
to have been remarked before. In particular, 
in~\cite{CDRSa} an infinite tower of
boundary states was suggested for a certain boundary condition
in the $a^{(1)}_2$ affine Toda theory,
based on the assumption that
such poles always correspond to new boundary bound states. While this
cannot be absolutely ruled out in the Toda context, it would now
appear to be an unnecessarily complicated scenario. Similar poles were
also discussed in the boundary sine-Gordon model in~\cite{SSa}, where
they were rejected on the basis of a Bethe ansatz solution of
a related lattice model. However, an interpretation within
the continuum boundary field theory was not given in that paper.

When $b$ is in the range $[1,2]$, there is also the possibility to
form the second boundary bound state as an excited state of the
first. Indeed, 
both $R^{[1]}_b$ and
$R^{[2]}_b$ have simple poles at $\theta=iV_{21}$, where
$V_{21}=(3{-}b)\pi/6$. 
These are labelled $C$ and $C'$ on figures~4 and~5, and the equality
\eq
e_2=e_1+M\cos V_{21}
\en
shows that their positions match such an interpretation.
We therefore deduce that the coupling $g_{21}$ is nonzero whenever
both $\Phi_1$ and $\Phi_2$ are in the spectrum.
The relevant on-shell diagrams are shown in figure~7.

The remaining poles are more delicate. Consider first the
continuation of the pole $C$ below $b=1$, labelled $D$ on figure~4.
Since $\Phi_2$ is not in the spectrum for $b<1$, 
figure~7 can no longer be invoked to explain this pole.
However, precisely as $b$ drops below $1$, it becomes geometrically
possible to draw a new on-shell diagram, as in figure~8.

{\begin{figure}[ht]
\[\begin{array}{cc}
{\epsfysize=0.4\linewidth\epsfbox{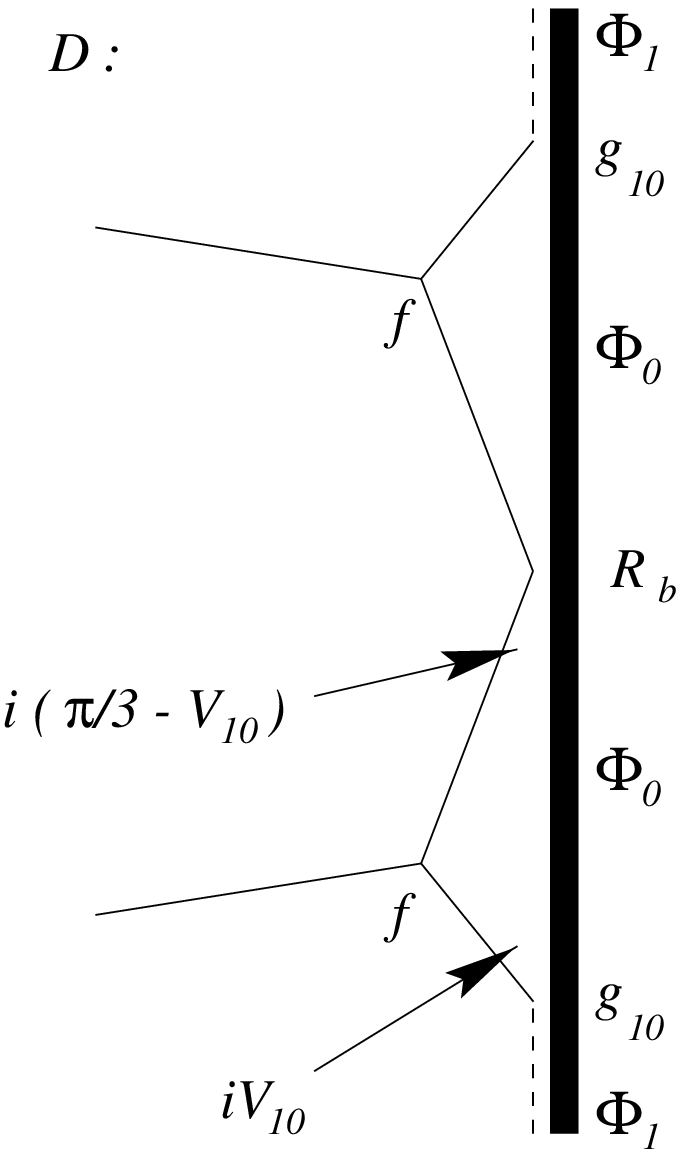}}
{}~~~~~~&~~~
{\epsfysize=0.4\linewidth\epsfbox{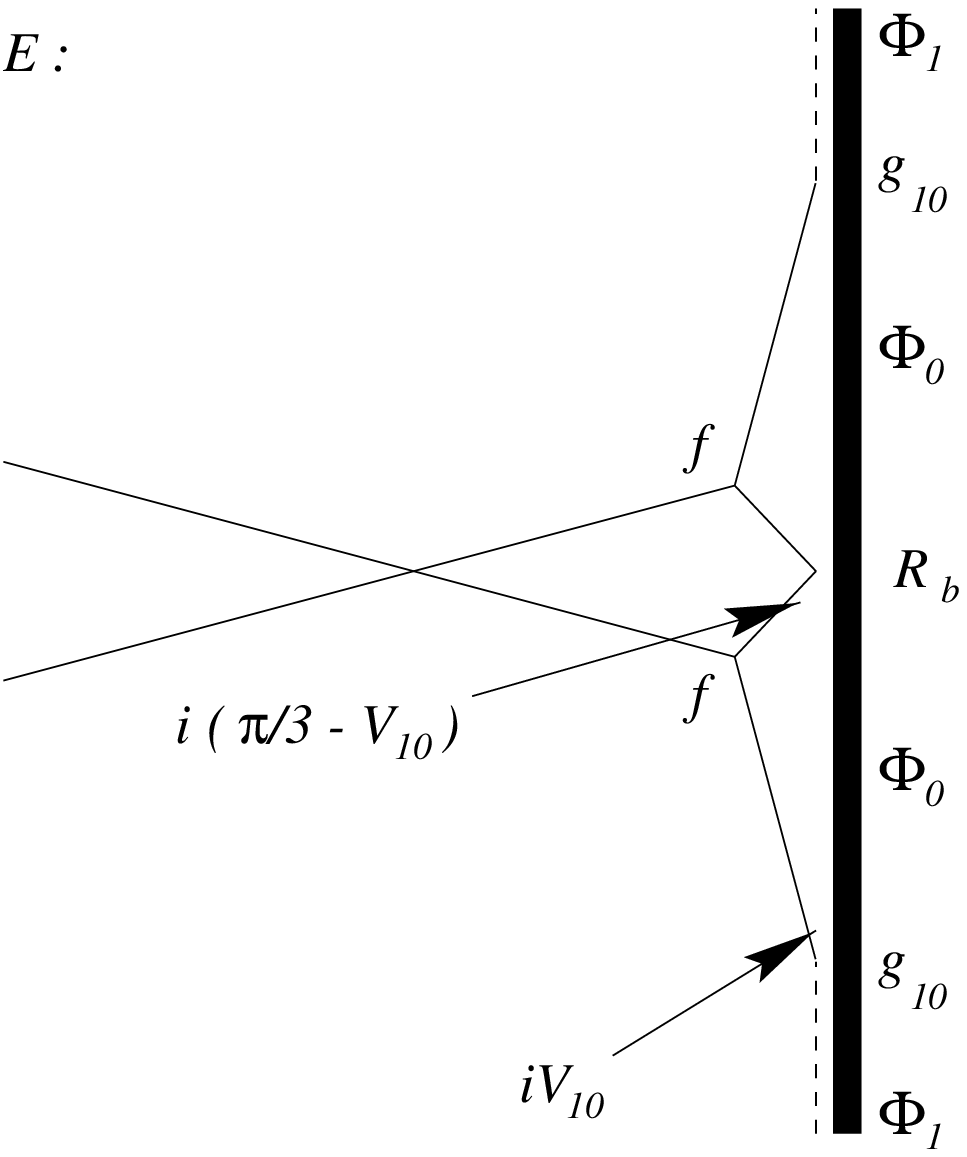}~~~{}}
\\[2pt]
\parbox[t]{.47\linewidth}{\small 8)
The pole $D$ in $R^{[1]}_b(\theta)$ 
at $i(2\pi/3-V_{20})$, present for $b\in [0,1]$.}
{}~~&~~
\parbox[t]{.47\linewidth}{\small 9)
The pole $E$ in $R^{[1]}_b(\theta)$ 
at $i(\pi/3+V_{21})$, present for $b\in [-1,0]$. }
\end{array}\]
\end{figure}}
 
\noindent
The quoted angles follow from the on-shell conditions at each
three-point vertex. The rapidity of the incoming particle is
$i(2\pi/3-V_{20})=i(3{-}b)\pi/6=iV_{21}$, which is indeed
the location of the pole $D$. However
the most straightforward application of the rule found in bulk
scattering would suggest that the diagram be associated with a
second-order pole, rather than the first order required for a match
with the exact reflection factor. This can be resolved by the same
generalisation of the Coleman-Thun mechanism as previously found to
operate in the bulk S-matrices of non self-dual affine Toda field
theories~\cite{CDSa}. Observe that the predicted higher pole should be
multiplied by a collection of vertex factors, and for the process
under discussion these comprise not only the three-point vertices
$f^2g_{\Onesmall_0}^2$, but also a ground-state reflection factor
$R_b(\theta)$, evaluated at $\theta=i(\pi/3-V_{10})=i(1{-}b)\pi/6$. 
But from (\ref{Rbdef}) or figure~2, $R_b(\theta)$ has a zero at
exactly this point. Thus the prediction is demoted to first order, as
required.

The remaining physical-strip poles are dealt with similarly. For $b$ in
the range $[-1,0]$, figure~8 can no longer be drawn, but in its place
the diagram drawn as figure~9 becomes possible. As before, a
generalised Coleman-Thun mechanism results in a first-order pole,
correctly-placed to match the pole marked $E$ on figure~4. Then
$R^{[2]}_b(\theta)$ has a double pole at $i(b{+}1)\pi/6$, marked $F$ on
figure~5. We leave it to the reader to check that two different
diagrams contribute. The first is similar to figure~8, but with the
sequence of boundary states down the right-hand side replaced by
$\{\Phi_2,\Phi_1,\Phi_1,\Phi_2\}$, while the second has the form of figure~9,
with the sequence  
$\{\Phi_2,\Phi_0,\Phi_0,\Phi_2\}$. In both the `internal' reflection
factor ($R^{[1]}_b$ or $R_b$ respectively) does not this time have a
zero at the relevant point, so the pole remains second-order. Finally,
there are various isolated values of $b$ at which higher-order poles
appear. It is straightforward to check that these also have
explanations in terms of `exceptional' on-shell diagrams, which can
only be drawn at these specific points.
%

%
%
\resection{Conclusions}
One perhaps unexpected bonus of the analysis just completed is that
it enables us to resolve a puzzle that was mentioned in the
concluding section of~\cite{Us1}. There it was noted that $R_{b=0}$,
the reflection factor for the $\Phi(h(0))$ boundary,
is exactly the same $R_{(1)}$, the reflection factor for the $\One$
boundary. This is in spite of the fact that the physics of the two
boundary conditions, as seen, for example, in finite-size spectra, are
very different. We now see that, while the functions are the same,
the pole structures have different interpretations in the two cases.
For $R_{(1)}(\theta)$, the pole at $i\pi/6$ can be traced back to the
on-shell diagram of figure~1a, and a boundary-particle coupling
for the $\One$ boundary equal to
$g_{\Onesmall_0}$. By contrast, equation~(\ref{ghform}) taken at
$b{=}0$ reveals that the corresponding coupling for the $\Phi(h(0))$
boundary, namely $g_0(h(0))$, is equal to $-g_{\Onesmall_0}$. 
Thus the contribution of the diagram of figure~1a in this case is
exactly the negative of that required for a match with the reflection
factor. Agreement is restored by the fact that
the residue at $i\pi/6$ picks up an exceptional extra contribution from
the formation of the $\Phi_1$ boundary bound state, the left hand
diagram of figure~3. Thus there are two consistent interpretations
of the $R_{(1)}$ reflection factor, depending on the sign given to the
bulk-boundary coupling. The `wrong' sign forces the introduction of a
boundary bound state to correct the residue at $i\pi/6$, but the
bootstrap can then be consistently closed on this extra state. A
complete set of infrared data in this situation therefore comprises not
only the bulk S-matrix and the elementary boundary reflection
factor, but also the sign of the bulk-boundary coupling. This sign
turns out to have an important role to play in the calculation of 
correlation functions through the form-factor approach, but we will
leave a detailed discussion of this point to another paper.

The restriction to real values of $b$ has confined the boundary field
$h$ to the range $[-|h_{\rm crit}|,|h_{\rm crit}|]$. What happens at 
other real values of $h$? For $h>|h_{\rm crit}|$, all seems to be well. 
Examining the formula~(\ref{fozzie}), we simply have to set 
$b=-3+i\hat b$, and continue away from $b=-3$ through real values 
$\hat b$. For such complex values of $b$, $R_b(\theta)$ is still
real-analytic. It has poles at $i\pi/6$ and $i\pi/2$ which can be
explained just as before, and a pair of physical-strip zeroes at
$i\pi/3\pm \hat b\pi/6$. These (or, better, their accompanying poles on
the unphysical sheet) might perhaps have an interpretation in terms of
unstable resonances, but in any case they are harmless as far as the
properties of the boundary reflection factor are concerned. By
contrast, for $h<-|h_{\rm crit}|$, the picture is less promising. The
continuation of (\ref{fozzie}) requires us to set $b=2+i\hat b$,
resulting in a reflection factor $R_b$ which is no longer real
analytic. In fact, such behaviour is only to be expected, given the 
observation of \cite{Us1} that values of the boundary field $h$ less
than $-|h_{\rm crit}|$ destabilise the bulk vacuum.

This completes the pole analysis of the model at all values of the
boundary field. We can claim to have employed a coherent set of rules, 
which has allowed us to rederive a pole structure and
spectrum that we knew independently (from BTCSA results) to
be correct. This should be useful in other contexts where BTCSA
results are not yet available. In particular it is possible to
resolve the discrepancies between boundary bootstrap and lattice Bethe
ansatz results for the boundary sine-Gordon model that were reported
in~\cite{SSa}. One aspect of this, the $u$-channel pole mechanism,
was mentioned above; a more complete analysis will appear
elsewhere. 

Another interesting feature revealed by the above analysis and our previous
paper~\cite{Us1} is the way that the BTBA equations 
rearrange themselves as the point $b{=}2$ is passed so as to 
continue to yield the correct ground state energy, even though the reflection
factors of the ground and first excited states themselves swap over.
Again, we expect such phenomena to reappear in more complicated models.

While the rules employed here have proved to be reliable, we should 
reiterate that they
have been obtained merely as the natural generalisation of the bulk
situation, rather than from any fundamental principles. Even to
verify some higher poles perturbatively would be reassuring, and if
the bulk~\cite{BCDSb} is anything to go by, the affine Toda field
theories may provide the most tractable examples. This might appear to
be a distant goal, given the complexities of the most basic
perturbative calculations in boundary models. However
the situation might simplify when attention is restricted to a check
of pole
residues, and the possibility should certainly be explored.

%
%

\bigskip
\bigskip
\noindent{\bf Acknowledgements --- }
We would like to thank Andrew Pocklington and Robert Weston for
discussions on related issues.
The work was supported in part by a TMR grant of the
European Commission, contract reference ERBFMRXCT960012,
and in part by an EPSRC grant
GR/K30667. PED and GMTW thank the organisers of the Oberwolfach
workshop on integrable systems for hospitality and the
EPSRC for Advanced Fellowships, and
RT thanks SPhT Saclay for hospitality.

%
%
\bigskip
%
\renewcommand\baselinestretch{0.95}

%
%
\end{document}